\documentclass[letterpaper]{article} 
\usepackage{aaai24}  
\usepackage{times}  
\usepackage{helvet}  
\usepackage{courier}  
\usepackage[hyphens]{url}  
\usepackage{graphicx} 
\usepackage{natbib}  
\usepackage{caption} 
\usepackage{algorithm}
\usepackage{algorithmic}

\usepackage{newfloat}
\usepackage{listings}
\DeclareCaptionStyle{ruled}{labelfont=normalfont,labelsep=colon,strut=off} 
\floatstyle{ruled}
\newfloat{listing}{tb}{lst}{}
\floatname{listing}{Listing}
\newcommand{\modelname}{Contra}
\usepackage{amsmath}
\usepackage{threeparttable}
\usepackage{multirow}
\usepackage{booktabs}
\usepackage{xcolor}
\usepackage{bbding}
\title{ContraNovo: A Contrastive Learning Approach to Enhance De Novo Peptide Sequencing}

\author{
  Zhi Jin\textsuperscript{\rm 1,2}\footnote{These authors contributed equally. The work was done during their internships at the Shanghai Artificial Intelligence Laboratory.},
  Sheng Xu\textsuperscript{\rm 3,1}$^{*}$,
  Xiang Zhang\textsuperscript{\rm 1,4}$^{*}$,
  Tianze Ling\textsuperscript{\rm 5},
  Nanqing Dong\textsuperscript{\rm 1},
  Wanli Ouyang\textsuperscript{\rm 1}\footnote{Corresponding author.},
  Zhiqiang Gao\textsuperscript{\rm 1}$^{\dag}$,
  Cheng Chang\textsuperscript{\rm 5}$^{\dag}$,
  Siqi Sun \textsuperscript{\rm 3,1}$^{\dag}$
}
\affiliations{
\textsuperscript{\rm 1} Shanghai Artificial Intelligence Laboratory,
\textsuperscript{\rm 2} Soochow University,
\textsuperscript{\rm 3} Fudan University,
\textsuperscript{\rm 4} University of British Columbia,\\
\textsuperscript{\rm 5} National Center for Protein Sciences (Beijing)\\
zjin@stu.suda.edu.cn, \{xusheng1,dongnanqing,ouyangwanli,gaozhiqiang,sunsiqi1\}@pjlab.org.cn, 
xzhang23@ualberta.ca,
ltz20@mails.tsinghua.edu.cn, changcheng@ncpsb.org.cn
}

\usepackage{bibentry}

\begin{document}

\maketitle

\begin{abstract}
De novo peptide sequencing from mass spectrometry (MS) data is a critical task in proteomics research. Traditional de novo algorithms have encountered a bottleneck in accuracy due to the inherent complexity of proteomics data. While deep learning-based methods have shown progress, they reduce the problem to a translation task, potentially overlooking critical nuances between spectra and peptides. In our research, we present ContraNovo, a pioneering algorithm that leverages contrastive learning to extract the relationship between spectra and peptides and incorporates the mass information into peptide decoding, aiming to address these intricacies more efficiently. Through rigorous evaluations on two benchmark datasets, ContraNovo consistently outshines contemporary state-of-the-art solutions, underscoring its promising potential in enhancing de novo peptide sequencing. The source code  is available at https://github.com/BEAM-Labs/ContraNovo.
\end{abstract}

\section{Introduction}

\begin{figure}[t]
\centering
\includegraphics[width=1\linewidth]{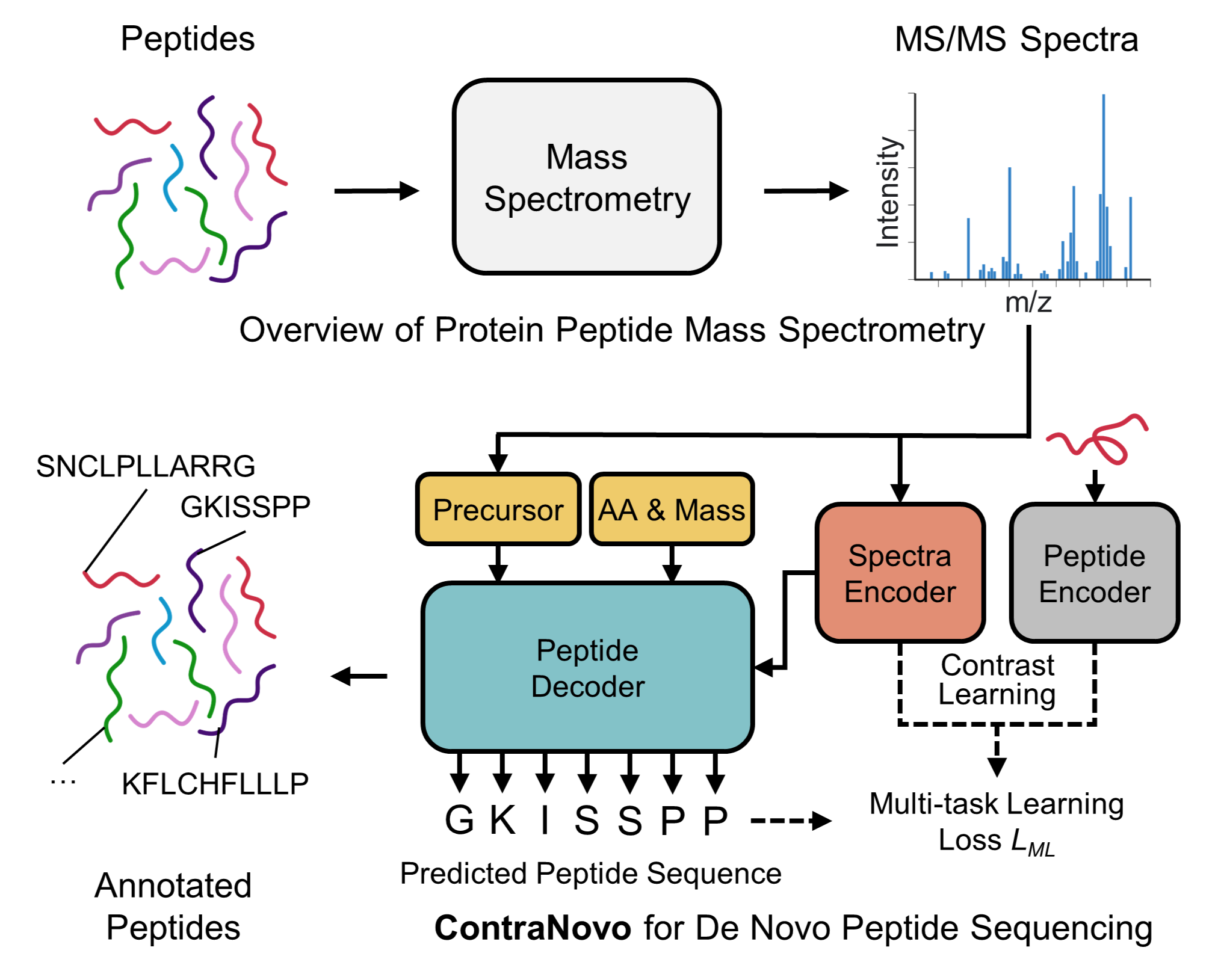}
\caption{De novo peptide sequencing process involves enzymatically cleaving peptides and analyzing them using mass spectrometry. They undergo purification and ionization, and selective fragmentation generates tandem mass spectra. Sequencing algorithms use this data to decode the peptide sequence. Deep learning-based algorithms can improve the accuracy of this process.}
\label{fig:seq}
\end{figure}

Peptide sequencing is an essential process for identifying peptide sequences from mass spectrum ~\cite{Aebersold2003b}. This technique has wide-ranging applications, particularly in the analysis of proteins from complex biological samples~\cite{chi2010pnovo}, the identification of potential drug targets for various diseases~\cite{lin2020relevant}, and the discovery of post-translational modifications on proteins~\cite{mann2003proteomic}. Database search-based algorithms have emerged as the predominant methods~\cite{ma2010challenges,noor2021mass}. These algorithms rank the peptide sequences stored in human-built databases based on the input spectrum, ultimately retrieving the most probable sequence. However, a significant limitation of these methods stems from the relatively limited number of peptide sequences available in the database compared to the vast array of peptides found in nature. As a result, the identification of peptides using such methods can be severely restricted.


With the emergence of deep learning techniques~\cite{lecun2015deep}, new approaches have surfaced to advance peptide mass spectrum sequencing. Particularly, deep neural networks have demonstrated remarkable generalization capabilities~\cite{novak2018sensitivity} , capturing underlying patterns that are common across diverse instances. This elevated level of generalizability successfully circumvents the restrictions of the traditional methodologies, enabling the generation of new peptides from scratch, a process known as de novo peptide sequencing, rather than relying solely on retrieving existing ones from databases. As a pioneering work in utilizing deep learning for de novo sequencing,~\citet{tran2017novo} developed a deep learning system named DeepNovo, which employed convolutional neural network (CNN) to learn spectrum features and utilized the long short-term memory (LSTM) architecture ~\cite{hochreiter1997long} for generating peptide sequences. This combination of CNN and LSTM exhibited promising results, paving the way for further advancements.

Recently, attention-based architectures~\cite{vaswani2017attention}, particularly the transformer model, have received considerable recognition for their exceptional results in an array of natural language processing tasks. Inspired by these accomplishments, \citet{yilmaz2022novo} proposed a transformer framework, CasaNovo, that exploits the self-attention mechanism to form a correlation between the sequence of mass spectral peaks and the sequence of amino acids. Their approach attained state-of-the-art results on a cross-species benchmark dataset, demonstrating the effectiveness of attention mechanisms in addressing this problem.

Nonetheless, existing models often oversimplify the de novo peptide sequencing challenge, regarding it only as a sequence-to-sequence task while disregarding the valuable connections between each spectrum and peptide. It's been observed that every spectrum holds plenty of information that can be utilized to infer specific groups of amino acids, and this information is frequently shared across different spectra of comparable peptides. Therefore, by contrasting the spectra of similar peptides, these shared features can be extracted more effectively, consequently aiding in the process of peptide generation.
Furthermore, the contrastive approach can also be utilized with spectra corresponding to unique peptides, enabling the model to differentiate between seemingly similar but ultimately distinct peptide sequences. In our approach, we use contrastive learning on a large set of training pairs to efficiently capture and leverage these pairwise connections. We call our method ContraNovo, which aims to improve the accuracy and robustness of de novo peptide sequencing by harnessing the power of contrastive learning.

Moreover, though mass spectrometry-based de novo sequencing algorithms heavily rely on the masses of amino acids and peptide fragments, previous methods haven't fully capitalized on the importance of this data. For instance, CasaNovo barely utilized amino acid masses during its training phase. To resolve these shortcomings, ContraNovo tackle this issue by incorporating amino acid masses in two unique ways.
First, we supply the decoder with essential inputs. These inputs include the inferred masses of already identified peptide fragments and temporarily uninferred masses. By incorporating this mass information, we enrich the context with valuable data to assist in the accurate generation of peptide sequences.
Second, we learn the amino acid feature representation utilizing the mass information. With the inclusion of mass data, the model is able to make better-informed decisions about the amino acid at the current position, leveraging the knowledge of mass compatibility.
ContraNovo effectively includes amino acid masses both during the decoding process. By doing so, it leverages this critical information to enhance the accuracy and efficacy of de novo peptide sequencing algorithms based on mass spectrometry.


\noindent In summary, our primary contributions to this field of study include:
\begin{itemize}
    \item We introduce ContraNovo, a precise de novo peptide sequencing method developed on contrastive learning. It significantly bridges the gap between mass spectral and peptide fragment features by implementing joint loss training.
    \item ContraNovo is the first method that incorporates the mass information of the prefix mass, suffix mass and amino acids into the decoding process of peptide sequencing algorithms.
    \item ContraNovo outperforms other methods and achieves the highest performance on two benchmark datasets. This firmly establishes it as a leading approach in the field.
\end{itemize}

\section{Related Work}
\noindent\textbf{Contrastive Learning.} Contrastive learning serves as a pivotal technique, empowering models to acquire meaningful data representations through the comparison and contrasting of diverse examples. In this approach, the model is trained to discern similarities and differences between pairs of data points, strategically bringing akin instances closer in a learned embedding space while simultaneously pushing apart dissimilar ones. This process unveils latent patterns and salient features within the data. Initially proposed by \citet{pmlr-v9-gutmann10a}, contrastive learning has found widespread utility in various tasks such as image representation learning~\cite{park2020contrastive}, text summarization~\cite{liu2022brio}, and object detection~\cite{xie2021detco}, among others. 

More recently, the concept of pretrained contrastive learning, exemplified by CLIP (Contrastive Language-Image Pretraining)~\cite{radford2021learning}, has significantly facilitated downstream tasks in image-text-related domains. Furthermore, CLIP's impact has extended into biological studies, where \citet{hong2021fastmsa} employed the contrastive loss on similar protein sequences to enhance multiple sequence alignment—a crucial aspect of proteomic research. \citet{altenburg2021yhydra}, on the other hand, applied the CLIP concept to mass spectrometry and peptides, leading to learned representations for both spectra and peptides. Distinguishing our work from the aforementioned research, we integrate contrastive learning into the learning process, culminating in substantial enhancements for generation of De Novo tasks. 

\noindent\textbf{Peptide De Novo Sequencing.}
The advancement of de novo peptide sequencing algorithms has been ongoing for over two decades~\cite{danvcik1999novo}. These methods range from traditional probabilistic models~\cite{ma2015novor} to computation models that leverage database searches~\cite{taylor1997sequence}, and up to advanced deep learning models that are widely used today. Notably, the use of transformer models has been significantly transformative in improved de novo peptide sequencing, as evidenced by recent studies. Pioneering research by \citet{yilmaz2022novo} has successfully applied transformer architectures to peptide sequence generation, yielding encouraging outcomes. Informed by these recent developments, we utilize an attention-based architecture complemented by multimodal strategies for the enhancement of the peptide sequence generation task. The fusion of these techniques is aimed at boosting the performance of de novo peptide sequencing, which underscores our contribution to the development of this ever-evolving research field.
 
\section{Methods}
\begin{figure*}[t]
\centering
\includegraphics[width=0.95\textwidth]{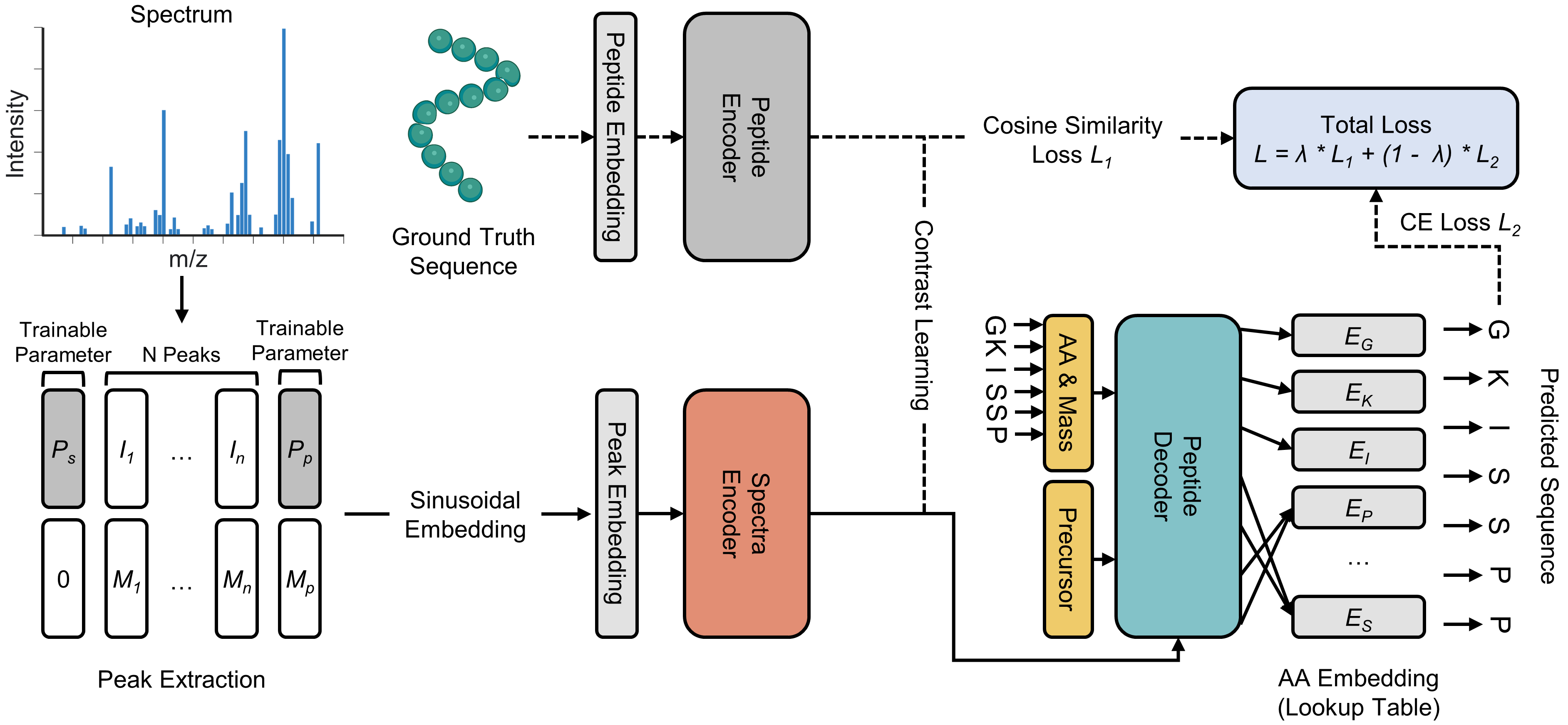}
\caption{The architecture of {\modelname}Novo is comprised of three main components: The red part emphasizes the feature extraction from the spectrum data, this extractor is trained by the contrastive learning loss.  The cyan part denotes the generation process with the peptide decoder, coupled with the  prefix-suffix mass information of leading amino acids. The output is then compared with the amino acid look up table. The light blue box is the total loss of the training strategy, which linearly combines the CE loss and CS loss. }
\label{fig:modelArchi}
\end{figure*}

{\modelname}Novo incorporates contrastive learning principles to amplify its spectrum feature extraction capability. By integrating the prefix mass, suffix mass, and the inherent attributes of {\modelname}Novo, a marked enhancement in the accuracy of de novo peptide sequencing is realized. The architectural of {\modelname}Novo is illustrated in Figure \ref{fig:modelArchi}. Subsequent sections delve into the detailed composition of its submodules.

\subsection{Contrastive Learning of {\modelname}Novo}

Mass spectrometry primarily produces m/z values and peak intensities, which are associated with the fragment ions of the peptides. Given this, it's rational to regard this data as a distinct modality compared to peptide sequences. Drawing inspiration from the CLIP~\cite{radford2021learning}, we sought to enhance de novo peptide sequencing accuracy through the contrastive learning of spectra and peptides. Our method demonstrated notable efficacy, as detailed below.

\noindent\textbf{Peak Embedding.} Our peak embedding methodology refers to the encoder approach from Casanovo, mirroring the positional encoder technique found in the transformer architecture. Specifically, for the m/z embedding, we employ an equal count of sine and cosine functions over a wavelength range from 0.001 to 10,000 m/z. The formula is given by:
\begin{equation}
\resizebox{0.91\hsize}{!}{$F_{emb} = \begin{cases}\sin((m/z)/(\frac{(M/Z)_{\max}}{(M/Z)_{\min}}(\frac{(M/Z)_{\min}}{2\pi})^{2i/d})),&\text{for}~i\leq d/2 \\\cos((m/z)/(\frac{(M/Z)_{\max}}{(M/Z)_{\min}}(\frac{(M/Z)_{\min}}{2\pi})^{2i/d})),&\text{for}~i > d/2\end{cases}$}
\end{equation}
where $(M/Z)_{\max}$ is 10,000, $(M/Z)_{\min}$ is 0.001, and \(d\) represents the peak embedding dimension, set at 512 in our implementation. 

It's crucial to note that using positional encoding empowers the model to adeptly discern the m/z disparities between peaks. These relative variations can directly inform on the peptide's amino acid composition. Building on this insight, we introduced zero peak and precursor peak embeddings. This addition aids the model in discerning the constitution of the initial and terminal amino acids. For flexibility, a trainable variable determines the intensities of both the zero and precursor peaks, with the m/z of the precursor peak being contingent upon the peptide mass.

\noindent\textbf{Spectrum and Peptide Encoder.} The encoders for both the spectrum and peptide employ the self-attention mechanism, a technique first popularized by the seminal work "Attention is All You Need" \cite{vaswani2017attention}. Since its introduction, self-attention has found utility across a plethora of domains for feature extraction. The core operation of self-attention is succinctly represented as:
\begin{equation}\label{eq:Attention}
\begin{split}
EnLayer(X) = softmax\left(\frac{(XW^Q)({XW^K})^T}{\sqrt{d_{k}}}\right)(XW^V)
\end{split}
\end{equation}
Here, ${d_{k_X}}$ stands for the dimension of the input feature, while matrices $W_X^Q$, $W_X^K$, and $W_X^V$ have dimensions of (${d_{k_X}} * {d_{k_X}}$).

\noindent\textbf{Global Feature Extraction.} The methodology for global feature extraction is illuminated by the attention mechanism. It employs a learnable vector parameter, termed the "global feature extractor," harmonized in dimensionality with the post-encoder feature. Utilizing the attention mechanism, we derive global feature vectors for both the spectrum and peptide. These vectors play a pivotal role in the ensuing contrastive learning process, expressed mathematically as:
\begin{equation}\label{eq:Global}
\begin{split}
Glo(X) & = softmax(V X^T)X
\end{split}
\end{equation}
Where \( V \) denotes the learnable vector parameter tasked with feature extraction.

\noindent\textbf{Contrastive Learning.} To establish a correlation between spectra and peptides, we employ a contrastive learning paradigm. This involves computing the cosine similarity, scaling pairwise cosine similarities, and subsequently determining the cross-entropy loss. The contrastive learning loss is captured by:
\begin{equation}
\begin{split}
CSLoss = - \left(\sum_{i=1}^{B_n}\sum_{j=1}^{B_n}\log\left(\frac{e^{CS_{ij}}}{\sum_{x=1}^{N} e^{CS_{xj}}}\right) * label_{ij} \right. \\ 
\left. +\sum_{i=1}^{B_n}\sum_{j=1}^{B_n}\log\left(\frac{e^{CS_{ij}}}{\sum_{x=1}^{B_n} e^{CS_{ix}}}\right) * label_{ij}\right) / 2
\end{split}
\end{equation}
Here, the $B_n$ is the batchsize and the ${label_{ij}}$ assumes a value of 1 when $i = j$ and 0 otherwise. ${CS_{ij}}$ represents the cosine similarity between spectra and peptide, as defined by:
\begin{align}
    CS_{ij} = \frac{Glo^{P_i}{\cdot}Glo^{S_j}}{\lvert Glo^{P_i} \rvert \lvert Glo^{S_j} \rvert}
\end{align}
with \(Glo^{P_i}\) and \(Glo^{S_i}\) denoting the global features of the peptide and spectrum, respectively. 

\subsection{Decoder of {\modelname}Novo}
The decoder architecture mirrors the foundational decoding scheme in the transformer model, but with adaptations catering to de novo peptide sequencing. Specifically, we incorporate both prefix and suffix sum masses of the peptide into the peptide embedding, enhancing amino acid inference. The peptide encoder further refines this process through its contribution to contrastive learning, thus optimizing decoding accuracy.

\noindent\textbf{Peptide Embedding.} This embedding comprises three components: index embedding, prefix mass sum embedding, and suffix mass sum embedding. The index embedding utilizes an embedding layer to represent the 20 amino acids and their modifications. For the embeddings of the prefix and suffix mass sums, we deploy the positional encoder elaborated upon earlier. 
The significance of prefix and suffix sums becomes evident when inferring the terminal amino acids of peptides. For example, by gauging the relative difference between peak m/z values and either the prefix or suffix mass sum, the model can readily decode specific amino acids. Illustratively, a zero suffix sum will direct the model end the process of peptide decoding.

\noindent\textbf{Peptide Decoder.} The peptide decoder integrates self-attention, encoder-decoder attention mechanisms, feed-forward networks, and layer normalization. This synergy produces an output sequence, grounded on previously generated tokens and encoder insights. 
The self-attention mechanism retains its original form in the decoder. However, within the encoder-decoder attention mechanism, the query vector \(Q_X\) is superseded by the peptide's encoded embedding. Notably, the decoder begins with the vector which concatenates three components: mass sum embedding with charge, a positional encoder with a prefix sum zero value, and a positional encoder with the precursor mass for the suffix sum.

\noindent\textbf{Amino Acid Embedding (Lookup Table).} For de novo peptide sequencing derived from spectral data, the amino acid mass, denoted as \(AA_{mass}\), is significant. Essentially, within the domain of spectrum-based proteomics, mass stands as the exclusive identifier for amino acids. To encapsulate this vital information, we deploy the embedding vector \(Concat(E(AA_{mass}), E(AA_{idx}))\). Here, \(E(AA_{mass})\) is the embedding associated with the amino acid mass, and \(E(AA_{idx})\) is the embedding of the amino acid index. Upon completion of training, these embedding vectors can be regarded as a lookup table. Through evaluating the cosine similarity between these embedding vectors and the decoder output, the corresponding amino acid can be discerned.

\subsection{Training with Joint Loss}
During the model's optimization phase, two objectives are pursued concurrently. The first aims to amplify the cosine similarity for contrastive learning, while the second strives to reduce the cross-entropy loss associated with the peptide decoder.

\noindent\textbf{Decoder Loss.} The loss function employed for the decoder is the cross entropy loss (CELoss). This loss function is widely recognized in classification problems and can be mathematically articulated as:
\begin{equation}
   CELoss(y, \hat{y}) = -\sum_{i=1}^{n} y_i \log(\hat{y}_i) 
\end{equation}
In this expression, \(y\) stands for the authentic probability distribution, and \(\hat{y}\) signifies the predicted distribution.

\noindent\textbf{Training Strategy}. To optimally use contrastive learning for de novo peptide sequencing, a combined training technique is adopted. The final loss function is represented as:
\begin{equation}
    Loss = (1 - \lambda)\ CELoss - \lambda \  CSLoss  
\end{equation}
The parameter \(\lambda\) is a dynamic hyperparameter in this context. If the $CSLoss$ is higher than 0.1, the $\lambda$ is 0.1, otherwise the $\lambda$ is 0.

\section{Experiments}

\subsection{Experimental Setup}

\noindent\textbf{Datasets.} Successful contrastive learning across varied modalities often demands a rich set of training data. In this context, CLIP~\cite{radford2021learning}, which has seen monumental success in domains such as image generation and recognition, was trained on a staggering 800 million image-text pairs. Inspired by this, we employed the expansive MassIVE-KB dataset~\cite{wang2018assembling} to bolster our contrastive learning representation. This dataset, earlier leveraged by studies like GLEAMS ~\cite{bittremieux2022learned} and CasaNovo, boasts 30 million high-resolution peptide-spectrum matches (PSMs), derived from a multitude of instruments, and encompassing numerous post-translational modifications.
For model validation and benchmarking against leading de novo peptide sequencing techniques, we employed the nine-species benchmark dataset presented by DeepNovo. This dataset collects approximately 1.5 million mass spectra from nine distinct experiments, all utilizing the same instrument but analyzing peptides from different species. Every spectrum is accompanied by a peptide sequence, confirmed through a database search identification at a standard false discovery rate (FDR) of 1\%.
In CasaNovo's latest iteration, the dataset underwent revision using the protein identification software Crux~\cite{McIlwain2014}, in alignment with a Percolator~\cite{Spivak2009} q-value \(< 0.01\), based on the same nine PRIDE datasets~\cite{martens2005pride}. This reformed dataset encapsulates seven variable modifications, including methionine oxidation, N-terminal acetylation, among others. Post-exclusion of shared peptides and PSMs across species, 2.8 million PSMs were accrued. For convenience, the revised dataset is termed 9-species-V2, while the original remains as 9-species-V1.

\noindent\textbf{Hyperparameters.} In this study, we mapped all inputs into a 512-dimensional vector space, which includes peaks, peptides, and amino acids. The peptide embedding specifically comprises 256 dimensions for the index of each amino acid, and another 256 dimensions for the prefix sum and suffix sum of amino acids. The amino acid embedding includes 256 dimensions for the index and 256 dimensions for the mass of each amino acid. The peptide encoder, spectrum encoder, and peptide decoder each employ 9 attention layers. All our attention layers come with 1024 feed forward dimensions.
During the training process, we used 8 A100 GPUs with a batch size of 4096. As a result, this setup created a contrastive learning cosine similarity matrix of dimensions 512*512 on each GPU. We set the learning rate at 0.0004 and applied a linear warm-up. For gradient updates, we used the AdamW optimizer~\cite{kingma2014adam}.

\noindent\textbf{Evaluation Metrics.} In our study, we used the evaluation metrics of CasaNovo~\cite{yilmaz2022novo} to assess the results on both amino acid (AA) and peptide levels.
At the AA level, we defined the amino acid is matched when (1) differ by $< 0.1$ Da in mass from the corresponding ground truth amino acid, and (2) have either a prefix or suffix that differs by no more than 0.5 Da in mass from the corresponding amino acid sequence in the ground truth peptide. Here, the precision at the AA level is represented as $N^a_{match}/N^a_{all}$, where $N^a_{all}$ is the total count of amino acids in the dataset, $N^a_{match}$ is the number of matched amino acids.
For the peptide level, the $N^{pep}_{match}$ signifies peptides where every amino acid matches the given dataset, the peptide precision is defined as $N^{pep}_{match} / N^{pep}_{all}$, where the $N^{pep}_{all}$ represents the total number of peptides in the dataset.
For a comparative analysis with CasaNovo, we employed a Precision-Coverage curve, which categorizes peptides based on their corresponding confidence scores. The confidence score is the mean probability of the constituent amino acids as identified by the model. The curve could indicate the prediction proficiency and stability of the model.


\noindent\textbf{Baselines.} To objectively evaluate the efficacy of our ContraNovo approach, we pitted it against several state-of-the-art tools, providing a broad perspective on its relative performance. The tools included:

\begin{itemize}
    \item \textbf{Peaks}~\cite{ma2003peaks}: This employs a protein database search strategy, which harnesses tandem mass spectra of tryptic peptides. By intertwining de novo sequencing with homology-based algorithms, Peaks aptly navigates the inherent ambiguities of peptide sequencing.
    
    \item \textbf{Deepnovo}~\cite{tran2017novo}: A deep learning-centric method, Deepnovo employs both convolutional neural networks and LSTM. Notably, it adopts the same cross-validation framework as CasaNovo.
    
    \item \textbf{PointNovo}~\cite{qiao2021computationally}: Built on the foundation of neural networks, PointNovo presents a de novo peptide sequencing model. Its distinct edge lies in its ability to seamlessly handle mass spectrometry data across varied resolution levels without escalating computational complexity.
    
    \item \textbf{CasaNovo}~\cite{yilmaz2022novo}: CasaNovo adopts a transformer-based architecture to achieve de novo peptide sequencing. It utilizes a leave-one-out cross-validation framework, where training occurs on data from eight species and testing on the remaining ninth species, iterating this for each of the nine species.
    
    \item \textbf{CasaNovoV2}~\cite{yilmaz2023sequence}: An evolved version of CasaNovo, CasaNovoV2 trains on the extensive MassIVE-KB dataset and integrates a beam search in its peptide decoding phase. Consequently, there's a marked upswing in sequencing accuracy.
\end{itemize}

\subsection{Results}
\begin{table*}[!htbp]
\setlength{\belowcaptionskip}{2mm}
\centering
\begin{threeparttable}
\setlength{\tabcolsep}{2mm}
\scalebox{0.95}
{
\begin{tabular}{l|cccccc|cccccc}
\toprule
&\multicolumn{6}{c}{Amino acid precision} & \multicolumn{6}{c}{Peptide precision}\\
Species & Peaks. & Deep. &  Point. & Casa. & Casa.V2 & {\modelname.}  & Peaks. & Deep. &  Point.  & Casa. & Casa.V2 & \modelname. \\
\midrule
Mouse         &0.600&0.623&0.626&0.689&0.760&\textbf{0.798}&0.197&0.286&0.355&0.426&0.483&\textbf{0.567}\\
Human         &0.639&0.610&0.606&0.586&0.676&\textbf{0.771}&0.277&0.293&0.351&0.341&0.446&\textbf{0.622}\\
Yeast         &0.748&0.750&0.779&0.684&0.752&\textbf{0.797}&0.428&0.462&0.534&0.490&0.599&\textbf{0.674}\\
M.mazei       &0.673&0.694&0.712&0.679&0.755&\textbf{0.799}&0.356&0.422&0.478&0.478&0.557&\textbf{0.630}\\
Honeybee      &0.633&0.630&0.644&0.629&0.706&\textbf{0.745}&0.287&0.330&0.396&0.406&0.493&\textbf{0.576}\\
Tomato        &0.728&0.731&0.733&0.721&0.785&\textbf{0.810}&0.403&0.454&0.513&0.521&0.618&\textbf{0.672}\\
Rice bean     &0.644&0.679&0.730&0.668&0.748&\textbf{0.807}&0.362&0.436&0.511&0.506&0.589&\textbf{0.677}\\
Bacillus      &0.719&0.742&0.768&0.749&0.790&\textbf{0.828}&0.387&0.449&0.518&0.537&0.622&\textbf{0.688}\\
Clam bacteria &0.586&0.602&0.589&0.603&0.681&\textbf{0.711}&0.203&0.253&0.298&0.330&0.446&\textbf{0.486}\\
\midrule
Average       &0.663&0.673&0.687&0.667&0.739&\textbf{0.785}&0.322&0.376&0.439&0.448&0.539&\textbf{0.621} \\
\bottomrule
\end{tabular}
}
\caption{Comparison of the performance of {\modelname}Novo and five baseline methods on 9-species-V1 test set. The bold font indicates the best performance.}\label{tab:testV1}
\end{threeparttable}
\end{table*}

As part of our commitment to highlighting the superiority of ContraNovo, we gauged its performance in contrast to many established baseline methods using two distinct peptide sequencing datasets. The evidence inarguably displays that ContraNovo delivers not only high sequencing accuracy but also exceptional generalization abilities. We conducted further ablation studies to corroborate the effectiveness of certain model components. In de novo peptide sequencing, the presence of amino acids of similar masses often poses challenges to accurate sequencing.  Our analysis demonstrate that ContraNovo has an exceptional performance in handling such cases.

\noindent\textbf{Performance of ContraNovo on 9-Species-V1 Benchmark Dataset.} The 9-species-V1 dataset stands as a cornerstone for assessing peptide sequencing algorithms, gaining prominence since DeepNovo's introduction. In this research, our main objective was to ascertain the prowess of ContraNovo in comparison to other prominent baselines. The detailed comparison is tabulated in Table \ref{tab:testV1}. ContraNovo outperforms all competitors, both at the amino acid level and peptide level, over all nine species.

\begin{figure}[th]
\centering
\includegraphics[width=\columnwidth]{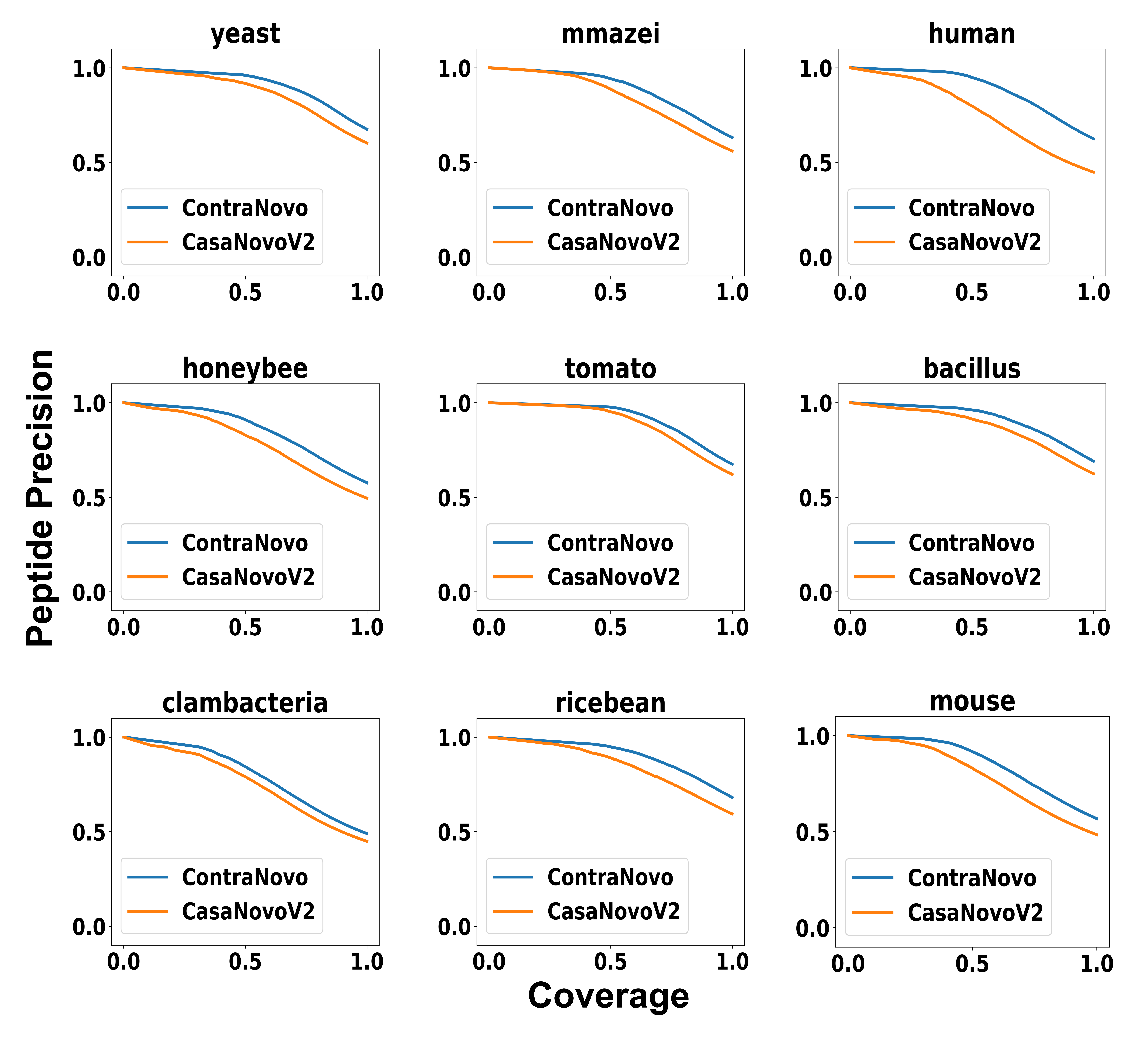}
\caption{The Peptide Precision-Coverage Curve. The horizontal axis represents coverage, and the vertical axis represents peptide recall. The blue line represents the performance of ContraNovo, and the yellow line represents the performance of CasaNovoV2. All of the blue lines in the image are positioned above the orange lines, demonstrating the advancement of ContraNovo.}
\label{fig:prc}
\end{figure}

Specifically, at the average amino acid level precision across all nine species, ContraNovo edges ahead of CasaNovoV2 by 4.6\%. Additionally, it beats CasaNovo, PointNovo, DeepNovo, and Peaks by margins of 11.8\%, 9.8\%, 11.4\%, and 12.2\% respectively. 
In the realm of proteomics, peptide-level accuracy is often of great interest. Here too, ContraNovo's superiority is evident. At the average peptide level precision across all nine species, ContraNovo surpasses CasaNovoV2, CasaNovo, PointNovo, DeepNovo, and Peaks by 8.2\%, 17.3\%, 18.2\%, 24.5\%, and 29.9\% respectively. A highlight of our study was ContraNovo's exceptional performance on human data, where it trumps CasaNovoV2 by a remarkable 17.6\%. These findings underscore ContraNovo's predictive prowess and broad-spectrum applicability. A visual representation of ContraNovo's exemplary performance is available in the Precision-Coverage curve, depicted in Figure \ref{fig:prc}. This graph is in alignment with our previous discussions, reinforcing the robust nature of ContraNovo's predictions.

\noindent\textbf{Performance of ContraNovo on 9-Species-V2 Benchmark Dataset.}
The 9-Species-V2 dataset, an improvement on the V1 dataset introduced by CasaNovoV2, includes a richer diversity of modified amino acids, thereby incrementing the sequencing complexity. To demonstrate ContraNovo's capacity to face this enhanced challenge, we conducted a stringent evaluation against this updated dataset.
As shown in Table \ref{tab:test_v2}, ContraNovo shows superior performance than CasaNovoV2. When we calculate average performances over the nine species datasets, ContraNovo exceeds CasaNovoV2 by 3.3\% in amino acid precision and 3.7\% in peptide precision. This high performance of ContraNovo is sustained across all species datasets. Minor discrepancies exist in the bacillus and yeast datasets; although ContraNovo doesn't markedly outperform CasaNovoV2 in these instances, it still yields a greater amino acid precision, and its peptide precision matches that of CasaNovoV2.
These findings not only attest to ContraNovo's skill in sequencing amidst increased modifications but also reflect its potent adaptability and generalization capabilities. Such traits endorse ContraNovo as an invaluable asset in handling complex and modified peptide sequencing datasets.

\begin{table}[!htbp]
\setlength{\belowcaptionskip}{2mm}
\centering
\begin{threeparttable}
\setlength{\tabcolsep}{2mm}
\scalebox{0.9}
{
\begin{tabular}{l|cc|cc}
\toprule
&\multicolumn{2}{c}{Amino acid precision} & \multicolumn{2}{c}{Peptide precision}\\
Species & Casa.V2 & Contra. & Casa.V2 & Contra. \\
\midrule
Mouse        &0.792&\textbf{0.816}&0.520&\textbf{0.566}\\
Human        &0.836&\textbf{0.893}&0.651&\textbf{0.745}\\
Yeast        &0.851&\textbf{0.875}&\textbf{0.729}&0.719\\
M.mazei      &0.828&\textbf{0.863}&0.675&\textbf{0.716}\\
Honeybee     &0.759&\textbf{0.804}&0.572&\textbf{0.608}\\
Tomato       &0.823&\textbf{0.860}&0.692&\textbf{0.716}\\
Rice bean    &0.858&\textbf{0.909}&0.702&\textbf{0.791}\\
Bacillus     &0.852&\textbf{0.861}&\textbf{0.717}&0.713\\
Clam bacteria&0.760&\textbf{0.778}&0.505&\textbf{0.521}\\
\midrule
Average & 0.818 & \textbf{0.851}& 0.640 &\textbf{0.677} \\ 
\bottomrule
\end{tabular}
}
\caption{Comparison of the performance of {\modelname}Novo and CasaNovoV2 methods on 9-species-V2 test set. The bold font indicates the best performance.}\label{tab:test_v2}
\end{threeparttable}
\end{table}

\noindent\textbf{Impact of Components on Performance.}
A comprehensive analysis was undertaken to highlight the significant roles of different components of ContraNovo. The ablation study shows the importance of four core components: beam search decoding, contrastive learning between spectra and peptides, amino acid embedding via a Lookup Table, and incorporation of prefix mass and suffix mass information. The results are presented in Table \ref{tab:ablation}.

Drawing inspiration from \cite{freitag2017beam}, we adopted beam search, a proven technique in de novo sequencing shown by its utilization in DeepNovo and CasaNovoV2. Following its implementation in ContraNovo, we witnessed improvement in sequence accuracy, with a 3.1\% increase at the amino acid level and a 2.8\% gain at the peptide level, compared to greedy decoding.
At the heart of ContraNovo's architecture is our key idea: that employing contrastive learning loss in sequencing and aligning it with amino acid embedding, along with integrating prefix mass and suffix mass information, can significantly boost de novo sequencing accuracy. We conducted unique training, intentionally excluding these elements and implemented a greedy decoding approach to confirm this theory.
Interestingly, the findings pointed out that ignoring the contrastive learning loss resulted in a decrease by 2.2\% at the amino acid level and 2.7\% at the peptide level. Further, exclusion of the amino acid embedding led to a fall of 2.1\% and 2.7\% at the amino acid and peptide levels, respectively. Neglecting data on prefix and suffix quality triggered a decline by 0.1\% and 3.1\%, respectively. These results underscore the critical role these modifications play in enhancing peptide sequencing accuracy.

\begin{table}[!htbp]
\setlength{\belowcaptionskip}{2mm}
\centering
\begin{threeparttable}
\setlength{\tabcolsep}{2mm}
\scalebox{1}
{
\begin{tabular}{cccc|c|c}
\toprule
\multirow{2}{*}{BS} & \multirow{2}{*}{CL} & \multirow{2}{*}{AA} & \multirow{2}{*}{Mass}  & Amino acid & Peptide\\
      & & & & Precision & Precision\\
\midrule
\Checkmark & \Checkmark & \Checkmark &\Checkmark & \textbf{0.828} & \textbf{0.688} \\
& \Checkmark & \Checkmark & \Checkmark  & 0.797 & 0.660\\
& & \Checkmark & \Checkmark & 0.775 & 0.633\\
 & \Checkmark& &\Checkmark &0.776 & 0.633\\
 & \Checkmark & \Checkmark & &  0.796 & 0.629\\
\bottomrule
\end{tabular}
}
\caption{Impact of components on performance for bacillus, \textbf{BS} represents the usage of beam search decoding. \textbf{CL} represents the training with contrastive learning loss. \textbf{AA} represents the usage with Amino acid embedding (Lookup Table). and \textbf{Mass} is the training/inference with the incorporation of prefix mass and suffix mass information. }
\label{tab:ablation}
\end{threeparttable}
\end{table}

\begin{figure}[th]
\centering
\includegraphics[width=\columnwidth]{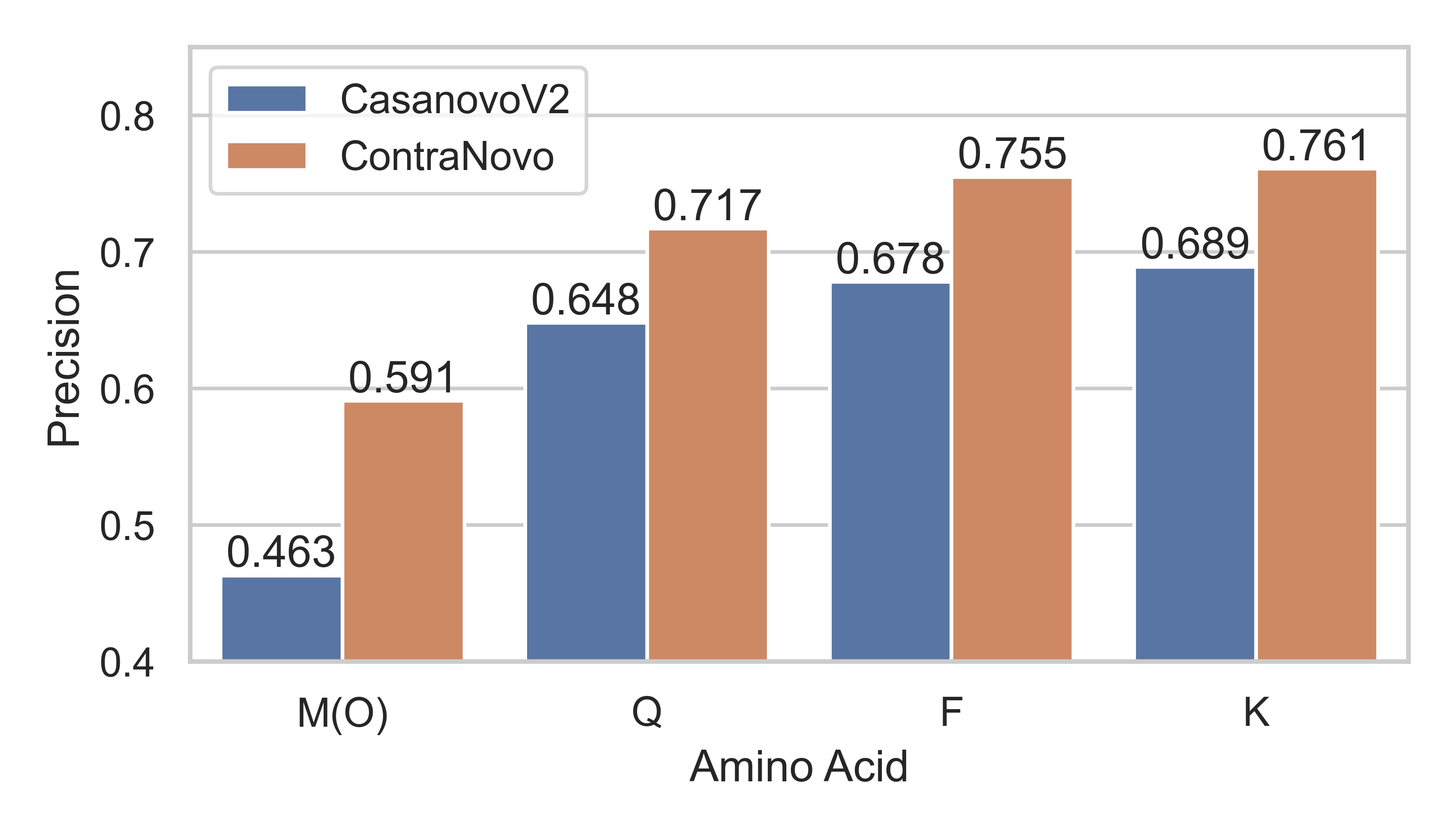}
\caption{The performance of predictions on similar-quality amino acids (F \& M(O) and Q \& K). The blue color represents CasaNovoV2, and the orange color represents ContraNovo. The orange bars in the graph are significantly higher than the blue bars, indicating that ContraNovo exhibits better performance in predicting similar-quality amino acids.}
\label{fig:case}
\end{figure}

\noindent\textbf{Performance of ContraNovo for Similar Mass Amino Acids.}
Peptide sequencing through mass spectrometry often grapples with challenges related to resolution. This is particularly significant given that the mass of amino acids is pivotal during sequencing. Several amino acids found in proteins bear strikingly similar masses. For instance, the mass difference between amino acids Q and K is a mere 0.46 Da, while F and M(Oxidation) differ by only 0.33 Da. Given the inherent resolution constraints of mass spectrometers, these minuscule differences can significantly influence the sequencing accuracy. 
To underscore the proficiency of ContraNovo in discerning amino acids with close masses, targeted tests were executed. Additionally, to juxtapose with CasaNovoV2, we used the checkpoint offered by CasaNovoV2 and gauged its accuracy on the entire 9-species-v1 dataset, focusing on amino acids of analogous mass. 
For a lucid comparative insight, the performances of ContraNovo and CasaNovoV2 are rendered as bar charts in Figure \ref{fig:case}.
The results lucidly indicate ContraNovo's superior accuracy over CasaNovoV2 in predicting these challenging amino acids. A standout observation is in the case of M(Oxidation), where ContraNovo eclipses CasaNovoV2 by an impressive margin of 13.8\%. This exemplar performance arguably cements ContraNovo's position as the avant-garde method in this domain.

\section{Conclusion}

In conclusion, our research presents ContraNovo, a noval approach in de novo peptide sequencing through contrastive learning. Our studies demonstrate an enhancement in peptide sequencing precision driven by the incorporation of contrastive learning and sums of mass prefixes and suffixes. Contrastive learning proves to be a potent force in connecting peptide and spectra data representations, with potential for further improvements on employing diverse datasets. The mutual reinforcement of the two tasks in our multi-task training method underscores its efficacy. As we navigate through the complex field of proteomics, ContraNovo stands as a significant contribution, equipping researchers with valuable insights and propelling further advancements in this ever-evolving domain.

\section{Acknowledgement}
This work is partially supported by the National Key R\&D Program of China (NO.2020YFE0202200, NO.2021YFA1301603, NO.2022ZD0160101) and the National Natural Science Foundation of China (NO.32088101). This work is also partially supported by the Focus Project of AI for Science of Comprehensive Prosperity Plan for Disciplines of Fudan University, Netmind.AI, and ProtagoLabs Inc.
\bibliography{aaai22}

\begin{thebibliography}{32}
\providecommand{\natexlab}[1]{#1}

\bibitem[{Aebersold and Mann(2003)}]{Aebersold2003b}
Aebersold, R.; and Mann, M. 2003.
\newblock {Mass spectrometry-based proteomics}.
\newblock \emph{Nature}, 422(6928): 198--207.

\bibitem[{Altenburg, Muth, and Renard(2021)}]{altenburg2021yhydra}
Altenburg, T.; Muth, T.; and Renard, B.~Y. 2021.
\newblock yhydra: Deep learning enables an ultra fast open search by jointly
  embedding ms/ms spectra and peptides of mass spectrometry-based proteomics.
\newblock \emph{bioRxiv}, 2021--12.

\bibitem[{Bittremieux et~al.(2022)Bittremieux, May, Bilmes, and
  Noble}]{bittremieux2022learned}
Bittremieux, W.; May, D.~H.; Bilmes, J.; and Noble, W.~S. 2022.
\newblock A learned embedding for efficient joint analysis of millions of mass
  spectra.
\newblock \emph{Nature methods}, 19(6): 675--678.

\bibitem[{Chi et~al.(2010)Chi, Sun, Yang, Song, Wang, Liu, Fu, Yuan, Wang, He
  et~al.}]{chi2010pnovo}
Chi, H.; Sun, R.-X.; Yang, B.; Song, C.-Q.; Wang, L.-H.; Liu, C.; Fu, Y.; Yuan,
  Z.-F.; Wang, H.-P.; He, S.-M.; et~al. 2010.
\newblock pNovo: de novo peptide sequencing and identification using HCD
  spectra.
\newblock \emph{Journal of proteome research}, 9(5): 2713--2724.

\bibitem[{Dan{\v{c}}{\'\i}k et~al.(1999)Dan{\v{c}}{\'\i}k, Addona, Clauser,
  Vath, and Pevzner}]{danvcik1999novo}
Dan{\v{c}}{\'\i}k, V.; Addona, T.~A.; Clauser, K.~R.; Vath, J.~E.; and Pevzner,
  P.~A. 1999.
\newblock De novo peptide sequencing via tandem mass spectrometry.
\newblock \emph{Journal of computational biology}, 6(3-4): 327--342.

\bibitem[{Freitag and Al-Onaizan(2017)}]{freitag2017beam}
Freitag, M.; and Al-Onaizan, Y. 2017.
\newblock Beam search strategies for neural machine translation.
\newblock \emph{arXiv preprint arXiv:1702.01806}.

\bibitem[{Gutmann and Hyvärinen(2010)}]{pmlr-v9-gutmann10a}
Gutmann, M.; and Hyvärinen, A. 2010.
\newblock Noise-contrastive estimation: A new estimation principle for
  unnormalized statistical models.
\newblock In Teh, Y.~W.; and Titterington, M., eds., \emph{Proceedings of the
  Thirteenth International Conference on Artificial Intelligence and
  Statistics}, volume~9 of \emph{Proceedings of Machine Learning Research},
  297--304. Chia Laguna Resort, Sardinia, Italy: PMLR.

\bibitem[{Hochreiter and Schmidhuber(1997)}]{hochreiter1997long}
Hochreiter, S.; and Schmidhuber, J. 1997.
\newblock Long short-term memory.
\newblock \emph{Neural computation}, 9(8): 1735--1780.

\bibitem[{Hong et~al.(2021)Hong, Sun, Zheng, Tan, and Li}]{hong2021fastmsa}
Hong, L.; Sun, S.; Zheng, L.; Tan, Q.; and Li, Y. 2021.
\newblock fastmsa: Accelerating multiple sequence alignment with dense
  retrieval on protein language.
\newblock \emph{bioRxiv}, 2021--12.

\bibitem[{Kingma and Ba(2014)}]{kingma2014adam}
Kingma, D.~P.; and Ba, J. 2014.
\newblock Adam: A method for stochastic optimization.
\newblock \emph{arXiv preprint arXiv:1412.6980}.

\bibitem[{LeCun, Bengio, and Hinton(2015)}]{lecun2015deep}
LeCun, Y.; Bengio, Y.; and Hinton, G. 2015.
\newblock Deep learning.
\newblock \emph{nature}, 521(7553): 436--444.

\bibitem[{Lin, Lin, and Lane(2020)}]{lin2020relevant}
Lin, E.; Lin, C.-H.; and Lane, H.-Y. 2020.
\newblock Relevant applications of generative adversarial networks in drug
  design and discovery: molecular de novo design, dimensionality reduction, and
  de novo peptide and protein design.
\newblock \emph{Molecules}, 25(14): 3250.

\bibitem[{Liu et~al.(2022)Liu, Liu, Radev, and Neubig}]{liu2022brio}
Liu, Y.; Liu, P.; Radev, D.; and Neubig, G. 2022.
\newblock BRIO: Bringing order to abstractive summarization.
\newblock \emph{arXiv preprint arXiv:2203.16804}.

\bibitem[{Ma(2010)}]{ma2010challenges}
Ma, B. 2010.
\newblock Challenges in computational analysis of mass spectrometry data for
  proteomics.
\newblock \emph{Journal of Computer Science and Technology}, 25(1): 107--123.

\bibitem[{Ma(2015)}]{ma2015novor}
Ma, B. 2015.
\newblock Novor: real-time peptide de novo sequencing software.
\newblock \emph{Journal of the American Society for Mass Spectrometry}, 26(11):
  1885--1894.

\bibitem[{Ma et~al.(2003)Ma, Zhang, Hendrie, Liang, Li, Doherty-Kirby, and
  Lajoie}]{ma2003peaks}
Ma, B.; Zhang, K.; Hendrie, C.; Liang, C.; Li, M.; Doherty-Kirby, A.; and
  Lajoie, G. 2003.
\newblock PEAKS: powerful software for peptide de novo sequencing by tandem
  mass spectrometry.
\newblock \emph{Rapid communications in mass spectrometry}, 17(20): 2337--2342.

\bibitem[{Mann and Jensen(2003)}]{mann2003proteomic}
Mann, M.; and Jensen, O.~N. 2003.
\newblock Proteomic analysis of post-translational modifications.
\newblock \emph{Nature biotechnology}, 21(3): 255--261.

\bibitem[{Martens et~al.(2005)Martens, Hermjakob, Jones, Adamski, Taylor,
  States, Gevaert, Vandekerckhove, and Apweiler}]{martens2005pride}
Martens, L.; Hermjakob, H.; Jones, P.; Adamski, M.; Taylor, C.; States, D.;
  Gevaert, K.; Vandekerckhove, J.; and Apweiler, R. 2005.
\newblock PRIDE: the proteomics identifications database.
\newblock \emph{Proteomics}, 5(13): 3537--3545.

\bibitem[{McIlwain et~al.(2014)McIlwain, Tamura, Kertesz-Farkas, Grant,
  Diament, Frewen, Howbert, Hoopmann, K{\"{a}}ll, Eng, MacCoss, and
  Noble}]{McIlwain2014}
McIlwain, S.; Tamura, K.; Kertesz-Farkas, A.; Grant, C.~E.; Diament, B.;
  Frewen, B.; Howbert, J.~J.; Hoopmann, M.~R.; K{\"{a}}ll, L.; Eng, J.~K.;
  MacCoss, M.~J.; and Noble, W.~S. 2014.
\newblock {Crux: Rapid Open Source Protein Tandem Mass Spectrometry Analysis}.
\newblock \emph{Journal of Proteome Research}, 13(10): 4488--4491.

\bibitem[{Noor et~al.(2021)Noor, Ahn, Baker, Ranganathan, and
  Mohamedali}]{noor2021mass}
Noor, Z.; Ahn, S.~B.; Baker, M.~S.; Ranganathan, S.; and Mohamedali, A. 2021.
\newblock Mass spectrometry--based protein identification in proteomics—a
  review.
\newblock \emph{Briefings in bioinformatics}, 22(2): 1620--1638.

\bibitem[{Novak et~al.(2018)Novak, Bahri, Abolafia, Pennington, and
  Sohl-Dickstein}]{novak2018sensitivity}
Novak, R.; Bahri, Y.; Abolafia, D.~A.; Pennington, J.; and Sohl-Dickstein, J.
  2018.
\newblock Sensitivity and generalization in neural networks: an empirical
  study.
\newblock \emph{arXiv preprint arXiv:1802.08760}.

\bibitem[{Park et~al.(2020)Park, Efros, Zhang, and Zhu}]{park2020contrastive}
Park, T.; Efros, A.~A.; Zhang, R.; and Zhu, J.-Y. 2020.
\newblock Contrastive learning for unpaired image-to-image translation.
\newblock In \emph{Computer Vision--ECCV 2020: 16th European Conference,
  Glasgow, UK, August 23--28, 2020, Proceedings, Part IX 16}, 319--345.
  Springer.

\bibitem[{Qiao et~al.(2021)Qiao, Tran, Xin, Chen, Li, Shan, and
  Ghodsi}]{qiao2021computationally}
Qiao, R.; Tran, N.~H.; Xin, L.; Chen, X.; Li, M.; Shan, B.; and Ghodsi, A.
  2021.
\newblock Computationally instrument-resolution-independent de novo peptide
  sequencing for high-resolution devices.
\newblock \emph{Nature Machine Intelligence}, 3(5): 420--425.

\bibitem[{Radford et~al.(2021)Radford, Kim, Hallacy, Ramesh, Goh, Agarwal,
  Sastry, Askell, Mishkin, Clark, Krueger, and Sutskever}]{radford2021learning}
Radford, A.; Kim, J.~W.; Hallacy, C.; Ramesh, A.; Goh, G.; Agarwal, S.; Sastry,
  G.; Askell, A.; Mishkin, P.; Clark, J.; Krueger, G.; and Sutskever, I. 2021.
\newblock Learning Transferable Visual Models From Natural Language
  Supervision.
\newblock arXiv:2103.00020.

\bibitem[{Spivak et~al.(2009)Spivak, Weston, Bottou, K{\"{a}}ll, and
  Noble}]{Spivak2009}
Spivak, M.; Weston, J.; Bottou, L.; K{\"{a}}ll, L.; and Noble, W.~S. 2009.
\newblock {Improvements to the Percolator Algorithm for Peptide Identification
  from Shotgun Proteomics Data Sets}.
\newblock \emph{Journal of Proteome Research}, 8(7): 3737--3745.

\bibitem[{Taylor and Johnson(1997)}]{taylor1997sequence}
Taylor, J.~A.; and Johnson, R.~S. 1997.
\newblock Sequence database searches via de novo peptide sequencing by tandem
  mass spectrometry.
\newblock \emph{Rapid communications in mass spectrometry}, 11(9): 1067--1075.

\bibitem[{Tran et~al.(2017)Tran, Zhang, Xin, Shan, and Li}]{tran2017novo}
Tran, N.~H.; Zhang, X.; Xin, L.; Shan, B.; and Li, M. 2017.
\newblock De novo peptide sequencing by deep learning.
\newblock \emph{Proceedings of the National Academy of Sciences}, 114(31):
  8247--8252.

\bibitem[{Vaswani et~al.(2017)Vaswani, Shazeer, Parmar, Uszkoreit, Jones,
  Gomez, Kaiser, and Polosukhin}]{vaswani2017attention}
Vaswani, A.; Shazeer, N.; Parmar, N.; Uszkoreit, J.; Jones, L.; Gomez, A.~N.;
  Kaiser, {\L}.; and Polosukhin, I. 2017.
\newblock Attention is all you need.
\newblock \emph{Advances in neural information processing systems}, 30.

\bibitem[{Wang et~al.(2018)Wang, Wang, Carver, Pullman, Cha, and
  Bandeira}]{wang2018assembling}
Wang, M.; Wang, J.; Carver, J.; Pullman, B.~S.; Cha, S.~W.; and Bandeira, N.
  2018.
\newblock Assembling the community-scale discoverable human proteome.
\newblock \emph{Cell systems}, 7(4): 412--421.

\bibitem[{Xie et~al.(2021)Xie, Ding, Wang, Zhan, Xu, Sun, Li, and
  Luo}]{xie2021detco}
Xie, E.; Ding, J.; Wang, W.; Zhan, X.; Xu, H.; Sun, P.; Li, Z.; and Luo, P.
  2021.
\newblock Detco: Unsupervised contrastive learning for object detection.
\newblock In \emph{Proceedings of the IEEE/CVF International Conference on
  Computer Vision}, 8392--8401.

\bibitem[{Yilmaz et~al.(2022)Yilmaz, Fondrie, Bittremieux, Oh, and
  Noble}]{yilmaz2022novo}
Yilmaz, M.; Fondrie, W.; Bittremieux, W.; Oh, S.; and Noble, W.~S. 2022.
\newblock De novo mass spectrometry peptide sequencing with a transformer
  model.
\newblock In \emph{International Conference on Machine Learning}, 25514--25522.
  PMLR.

\bibitem[{Yilmaz et~al.(2023)Yilmaz, Fondrie, Bittremieux, Nelson, Ananth, Oh,
  and Noble}]{yilmaz2023sequence}
Yilmaz, M.; Fondrie, W.~E.; Bittremieux, W.; Nelson, R.; Ananth, V.; Oh, S.;
  and Noble, W.~S. 2023.
\newblock Sequence-to-sequence translation from mass spectra to peptides with a
  transformer model.
\newblock \emph{Biorxiv}, 2023--01.

\end{thebibliography}
\end{document}